\documentclass[nofootinbib,prd,aps,superscriptaddress,preprintnumbers]{revtex4}
\pdfoutput=1
\usepackage{amsmath,amssymb,euscript}
\usepackage{slashed}
\usepackage{color}
\usepackage{accents}
\usepackage{hyperref}
\usepackage{ulem}
\usepackage{epsfig}
\usepackage{varioref}
\usepackage{xcolor}
\usepackage{verbatim}

\def\beq{\begin{equation}}
\def\eeq{\end{equation}}

\renewcommand{\emph}{\textit}

\graphicspath{{figs/}}

\begin{document}

\begin{flushright}

\end{flushright}

\title{Long live the Higgs portal!}

\author{Anastasiia Filimonova, Susanne Westhoff\\
\textit{ \small{Institute for Theoretical Physics, Heidelberg University, D-69120 Heidelberg, Germany} }}

\vspace{1.0cm}
\begin{abstract}
\vspace{0.2cm}\noindent
In Higgs portal models of fermion dark matter, scalar couplings are unavoidably suppressed by strong bounds from direct detection experiments. As a consequence, thermal dark matter relics must coexist with mediators in a compressed spectrum of dark particles. Small couplings and small mass splittings lead to slow mediator decays, leaving signatures with displaced vertices or disappearing tracks at colliders. We perform a comprehensive analysis of long-lived mediators at the LHC in the context of a minimal dark matter model with a naturally small Higgs portal, also known as the wino-bino scenario in supersymmetry. Existing searches for disappearing charged tracks and displaced hard leptons already exclude tiny portal couplings that cannot be probed by current direct and indirect detection experiments. For larger portal couplings, we predict new signatures with displaced soft leptons, which are accessible with run-II data. Searches for displaced particles are sensitive to weakly coupling mediators with masses up to the TeV scale, well beyond the reach of prompt signals.
\end{abstract}
\maketitle
\tableofcontents
\section{Introduction}
\noindent
The hypothesis of thermal Higgs portal dark matter had almost been declared a dead end. If dark matter interacts with the standard model through the Higgs boson, the portal strength is indeed strongly constrained by null results from direct detection experiments~\cite{Kim:2006af,Essig:2007az,Fedderke:2014wda,Beniwal:2015sdl,Escudero:2016gzx,Arcadi:2017kky,Athron:2018hpc}. A dark matter relic from thermal freeze-out would thus be overabundant today, unless its annihilation rate in the early universe was enhanced, for instance, by annihilation through a resonance or co-annihilation with mediators, or altered due to a non-standard cosmological history. In Higgs-portal scenarios with fermion dark matter, mediators are a necessary prediction of a UV-complete theory. In this work, we assume that dark matter is a fermion with no charges under the standard model gauge group, interacting only through electroweak mixing. Fermionic mediators with renormalizable Higgs couplings induce generally strong dark matter-nucleon scattering through electroweak mixing~\cite{Mahbubani:2005pt,Dedes:2014hga,Freitas:2015hsa,Tait:2016qbg,Lopez-Honorez:2017ora}. Mediator triplets or higher multiplets of the weak gauge group lead to non-renormalizable Higgs couplings, which are naturally suppressed by a high cutoff scale~\cite{Bharucha:2017ltz}. The small portal coupling implies slow decays of the heavy dark states. This can result in a decay length of several centimeters or more, before the mediator decays into dark matter and leptons or hadrons. At the LHC, long-lived mediators thus leave characteristic signatures of displaced vertices or disappearing charged tracks in the detector.

Higgs-portal scenarios with small dark matter couplings are thus a perfectly viable option for thermal dark matter that can be tested at colliders. In this work, we explore the prospects to discover thermal Higgs-portal dark matter through signatures with long-lived mediators at the LHC. We focus on a non-renormalizable Higgs portal with a Majorana fermion singlet and a fermion triplet in the adjoint representation of the weak gauge group. The portal coupling is naturally small and thus evades direct detection bounds. Our model can be UV-completed for instance by a fermion doublet or a scalar triplet. The former construction is known as the wino-bino scenario in the context of supersymmetry~\cite{ArkaniHamed:2006mb}, the latter is similar to type-II seesaw models for neutrino masses~\cite{Ma:2008cu,Choubey:2017yyn}. Due to its rich dark matter phenomenology and promising collider signatures, the small coupling regime of this and similar models has recently received increased interest~\cite{Lee:2018pag,Curtin:2018mvb,Rolbiecki:2015gsa,Nagata:2015pra,Bharucha:2018pfu,Calibbi:2018fqf}. We perform a comprehensive analysis of two scenarios: a scalar and a pseudo-scalar Higgs portal. While the scalar scenario has been the focus of most previous collider studies, the phenomenology of long-lived mediators in the pseudo-scalar scenario is still largely unexplored. We point out the characteristic differences between both scenarios, regarding their dark matter and collider phenomenology. We propose new signatures with displaced particles and show that they can probe mediator masses well beyond the reach of prompt signatures at the LHC.

This article is organized as follows. In Section~\ref{sec:model}, we introduce our model and discuss the relations between the scalar and pseudo-scalar Higgs portals. Section~\ref{sec:susy} is a brief interlude on the interpretation of our model in the minimal supersymmetric standard model (MSSM). In Section~\ref{sec:scattering}, we derive constraints on the portal strength from direct detection experiments. Section~\ref{sec:annihilation} is devoted to a detailed discussion of the relic dark matter abundance in different regimes of a small portal coupling. In Section~\ref{sec:lhc}, we eventually explore the collider phenomenology of the two scenarios under the assumption of a thermal dark matter candidate. We conclude in Section~\ref{sec:conclusions}.

\section{The singlet-triplet Higgs portal}\label{sec:model}
\noindent
We extend the standard model by two self-adjoint fermion fields with vector-like weak interactions,
\begin{align}
\chi_S = \chi_S^0,\qquad \chi_T = \begin{pmatrix}
\chi_T^0/\sqrt{2} & \chi^+ \\ \chi^{-} & -\chi_T^0/\sqrt{2}\end{pmatrix}.
\end{align}
Here $\chi_S$ is a standard-model singlet Majorana fermion and $\chi_T$ transforms under the weak gauge group as a triplet with zero hypercharge. We assume a discrete $\mathbb{Z}_2$ symmetry, under which $\chi_S$ and $\chi_T$ are odd and all standard-model particles are even. The lightest fermion state is stable and a dark matter candidate. In this scenario, there are no renormalizable Higgs couplings to dark fermions. At energies below a cutoff scale $\Lambda$, the scalar sector is described by the effective Lagrangian~\footnote{We use four-component notation for the fermion spinors.}
\begin{align}\label{eq:lag}
\mathcal{L}_{\text{eff}} = & - \frac{m_S}{2} \overline{\chi}_S\chi_S + \frac{1}{2}\,\overline{\chi}_S i\slashed{\partial}\,\chi_S - \frac{m_T}{2} \text{Tr}[\overline{\chi}_T\chi_T] + \frac{1}{2}\,\text{Tr}[\overline{\chi}_T i \slashed{D}\,\chi_T]\\\nonumber
& + \frac{\kappa_{ST}}{\Lambda}\Big((H^\dagger \overline{\chi}_T H)\chi_S +  \overline{\chi}_S (H^\dagger \chi_T H) \Big)\\\nonumber
&  - \frac{\kappa_S}{\Lambda}H^\dagger H\,\overline{\chi}_S\chi_S - \frac{\kappa_T}{\Lambda} H^\dagger H\,\text{Tr}[\overline{\chi}_T\chi_T] + \frac{\kappa_T'}{\Lambda}\Big(H^\dagger \overline{\chi}_T \chi_T H - \text{Tr}[\overline{\chi}_T H H^\dagger \chi_T]\Big),
\end{align}
where $D_\mu = \partial_\mu - i g\,\tau^a W_\mu^a$ is the covariant derivative, $\tau^a$ are the generators of the weak SU(2) gauge group, and $H = (h^+,(v+h+i\eta)/\sqrt{2})^\top$ is the Higgs doublet of the standard model. Gauge-invariant contractions of the fields are implicitly assumed. Higgs couplings to dark fermions are of mass dimension five and thus naturally small at energies well below $\Lambda$. Since we will focus on the parameter region with a high cutoff scale $\Lambda \gg M_W$, the impact of operators with higher mass dimensions can be neglected. We assume all parameters to be real in order to preserve CP invariance. The dark matter and collider phenomenology of the Higgs-portal couplings $\kappa_S$ and $\kappa_T$ has been extensively studied~\cite{LopezHonorez:2012kv,Djouadi:2011aa,Cirelli:2014dsa,Baum:2017enm}. In general, a complete theory that induces a singlet-triplet Higgs portal can also generate singlet-singlet and/or triplet-triplet Higgs couplings. However, since $\kappa_S$ and $\kappa_T,\,\kappa_T'$ will play essentially no role in our phenomenological analysis, we neglect them by setting $\kappa_S = \kappa_T = \kappa_T' = 0$.\\

After electroweak symmetry breaking, $\chi_S^0$ and $\chi_T^0$ mix through the Higgs portal. The scalar Lagrangian for the neutral fermions reads
\begin{align}\label{eq:lag-mass}
\mathcal{L}_S =  -\frac{m_T}{2} \overline{\chi}_T^0\chi_T^0 - \frac{m_S}{2} \overline{\chi}_S^0\chi_S^0 - \frac{\mu}{2}\Big(1 + \frac{2h}{v} + \frac{h^2}{v^2}\Big)\left(\overline{\chi}_T^0\chi_S^0 + \overline{\chi}_S^0\chi_T^0\right),\qquad \mu = \frac{\kappa_{ST} v^2}{\sqrt{2}\Lambda}.
\end{align}
We can choose $m_T$ positive without losing generality. The sign of $\mu$ is not observable in tree-level processes, so we assume it to be positive as well. Furthermore, we always request $m_T > - m_S$, so that the singlet fermion is lighter than the triplet. Due to electroweak mixing, the gauge eigenstates are not mass eigenstates of our model. We define a mass matrix $M$ by
\begin{align}
\mathcal{L} = - \frac{1}{2} M_{ij} \overline{\psi}_i\psi_j,\quad
M = \begin{pmatrix}
m_S & \mu\\
\mu & m_T
\end{pmatrix},\quad \psi = \begin{pmatrix}
\chi_S^0\\
\chi_T^0
\end{pmatrix}.
\end{align}
By diagonalizing $M$, the physical eigenstates $\chi_\ell$ and $\chi_h$ are readily obtained as
\begin{align}
\begin{pmatrix}
\chi_\ell\\
\chi_h
\end{pmatrix}
= \begin{pmatrix}
 \cos\theta\,\chi_S^0 - \sin\theta\,\chi_T^0\\
 \sin\theta\,\chi_S^0 + \cos\theta\,\chi_T^0
\end{pmatrix},\qquad \tan(2\theta) = \frac{2\mu}{m_T - m_S},\qquad 0 < \theta < \frac{\pi}{4}.
\end{align}
The corresponding mass eigenvalues for the neutral states $\chi_\ell$, $\chi_h$ and the charged states $\chi^\pm$ are
\begin{align}
m_{h,\ell} = \frac{1}{2}\Big(m_T + m_S \pm \Delta m_{h\ell}\Big),\qquad \Delta m_{h\ell} = \sqrt{(m_T-m_S)^2+4\mu^2},\qquad m_c = m_T.
\end{align}
For $m_S > \mu^2/m_T$, the mass $m_\ell$ of the lightest state is positive. The physical Lagrangian then reads (neglecting interactions with two Higgs bosons)
\begin{align}\label{eq:lag-scalar}
\widehat{\mathcal{L}}_S = & - \frac{m_\ell}{2}\,\overline{\chi}_\ell\chi_\ell - \frac{m_h}{2}\,\overline{\chi}_h\chi_h - \frac{\mu}{v} h \Big[\cos(2\theta) \big(\overline{\chi}_\ell \chi_h + \overline{\chi}_h \chi_\ell\big) + \sin(2\theta) \big(-\overline{\chi}_\ell \chi_\ell + \overline{\chi}_h \chi_h\big)\Big]\\\nonumber
& + g\,W_\mu^+\Big[- \sin\theta \left( \overline{\chi}_\ell \gamma^\mu \chi^- - \overline{\chi}^-\gamma^\mu \chi_\ell \right) + \cos\theta \left(\overline{\chi}_h \gamma^\mu \chi^- - \overline{\chi}^- \gamma^\mu \chi_h \right) \Big] + h.c.\\\nonumber
& + g\left(\cos\theta_w Z_\mu - \sin\theta_w A_\mu \right) \big[ \overline{\chi}^-\gamma^\mu \chi^+ - \overline{\chi}^+\gamma^\mu \chi^- \big],
\end{align}
where $\theta_w$ denotes the weak mixing angle. Singlet-triplet mixing induces scalar Higgs couplings and charged vector currents of the lightest state $\chi_\ell$. Neutral currents of $\chi_\ell$ and $\chi_h$ are absent due to the Majorana nature of the neutral fermions. In the limit of small mixing, the lightest state is mostly a weak singlet with suppressed couplings to the standard model.\\

A pseudo-scalar singlet-triplet Higgs portal can be obtained through a chiral rotation of the singlet fermion,
\begin{align}
\chi_S \to \text{exp}(i \gamma_5 \pi/2)\,\chi_S.
\end{align}
By applying this transformation to the Lagrangian in Eq.~(\ref{eq:lag-mass}), we obtain
\begin{align}\label{eq:lag-mass-p}
\mathcal{L}_P = - \frac{m_T}{2} \overline{\chi}_T^0\chi_T^0 + \frac{m_S}{2} \overline{\chi}_S^0\chi_S^0 - \frac{\mu}{2}\Big(1 + \frac{2h}{v} + \frac{h^2}{v^2}\Big)\left(\overline{\chi}_T^0 i\gamma_5\chi_S^0 + \overline{\chi}_S^0 i\gamma_5\chi_T^0\right).
\end{align}The chiral rotation turns the scalar portal into a pseudo-scalar portal and flips the sign of $m_S$. In the mass eigenbasis, we obtain
\begin{align}\label{eq:estates-p}
\begin{pmatrix}
\chi_\ell\\
\chi_h
\end{pmatrix}
= \begin{pmatrix}
 \cos\theta\,\chi_S^0 + \sin\theta\,i\gamma_5\chi_T^0\\
  \sin\theta\,i \gamma_5 \chi_S^0 + \cos\theta\,\chi_T^0
\end{pmatrix},\qquad m_{h,\ell}' = \frac{1}{2}\Big( \Delta m_{h\ell}' \pm (m_T + m_S)\Big) = \pm\,m_{h,\ell}.
\end{align}
The physical mass terms and interactions are given by
\begin{align}\label{eq:lag-pseudo-scalar}
\widehat{\mathcal{L}}_P= & - \frac{m_{\ell}'}{2}\,\overline{\chi}_\ell\chi_\ell - \frac{m_{h}'}{2}\,\overline{\chi}_h\chi_h - \frac{\mu}{v} h \Big[\cos(2\theta) \big(\overline{\chi}_\ell i\gamma_5 \chi_h + \overline{\chi}_h i\gamma_5 \chi_\ell\big) + \sin(2\theta) \big(\overline{\chi}_\ell \chi_\ell + \overline{\chi}_h \chi_h\big)\Big]\\\nonumber
& + g\,W_\mu^+\Big[ \sin\theta \left(\overline{\chi}_\ell\,i \gamma^\mu\gamma_5 \chi^- + \overline{\chi}^- i \gamma^\mu\gamma_5 \chi_\ell \right) + \cos\theta \left(\overline{\chi}_h \gamma^\mu \chi^- - \overline{\chi}^- \gamma^\mu \chi_h \right) \Big] + h.c.\\\nonumber
& + g\left(\cos\theta_w Z_\mu - \sin\theta_w A_\mu \right) \big[ \overline{\chi}^-\gamma^\mu \chi^+ - \overline{\chi}^+\gamma^\mu \chi^- \big].
\end{align}
Now the Higgs interactions with one heavy and one light neutral fermion are pseudo-scalar, while heavy-heavy and light-light couplings remain scalar. The gauge couplings of $\chi_\ell$ have an axial-vector structure.

\begin{figure}[!t]
\centering
\includegraphics[height=0.9in]{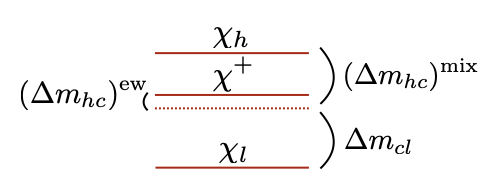} \hspace*{1.5cm} \includegraphics[height=0.9in]{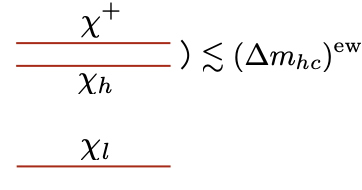}
\vspace*{0.2cm}
\caption{\label{fig:spectrum} Spectrum of dark fermions mixing through a Higgs portal. Left: ``normal'' mass hierarchy for sizeable portal coupling. Right: ``inverted'' hierarchy for small portal coupling.}
\end{figure}

The chiral rotation also flips the sign of the lightest mass eigenvalue in the spectrum, see Eq.~(\ref{eq:estates-p}). The parameter space $m_S < \mu^2/m_T$ with negative mass $m_\ell < 0$ in the scalar Lagrangian $\widehat{\mathcal{L}}_S$ thus corresponds to positive mass $m_{\ell}' > 0$ in the pseudo-scalar Lagrangian $\widehat{\mathcal{L}}_P$. We identify two physical scenarios,
\begin{align}\label{eq:scenarios}
scalar\,scenario: & \quad \widehat{\mathcal{L}}_S \text{ with } m_S > \mu^2/m_T\ \leftrightarrow\ m_{\ell} > 0,\qquad \qquad \\\nonumber
pseudo\text{\,-}scalar\,scenario: & \quad \widehat{\mathcal{L}}_P \text{ with } m_S < \mu^2/m_T\ \leftrightarrow\ m_{\ell}' > 0.
\end{align}
It is instructive to study the features of our model in the limit of small singlet-triplet mixing. In this limit, the couplings of the lightest state $\chi_\ell$ are approximated by
\begin{align}\label{eq:couplings-small}
W^+_\mu \overline{\chi}_\ell\,\gamma^\mu(i\gamma_5) \chi^- \sim g\,\theta,\quad h\,\overline{\chi}_\ell \chi_\ell \sim \frac{\mu}{v}\theta,\quad h\,\overline{\chi}_\ell\, (i\gamma_5) \chi_h \sim \frac{\mu}{2v};\qquad \theta \approx \frac{\mu}{m_T - m_S} \ll \frac{\pi}{4}.
\end{align}
For fixed values of $m_T$ and $\mu$, the mixing $\theta$ in the pseudo-scalar scenario is smaller than in the scalar scenario, see Eq.~(\ref{eq:scenarios}). Gauge and diagonal Higgs couplings of the lightest state are thus weaker in the pseudo-scalar scenario. The masses of the neutral fermions are approximated as
\begin{align}\label{eq:spectrum-small}
m_h = m_h' \approx m_T + \theta \mu,\qquad m_\ell = -m_\ell' \approx m_S - \theta \mu.
\end{align}
The lightest state $\chi_\ell$ is thus mostly a gauge singlet, while the heavier states $\chi^+$ and $\chi_h$ approximately correspond to the charged and neutral components of a weak triplet. In both scenarios, the mass splitting between the heavier states is given by
\begin{align}\label{eq:deltahc}
\Delta m_{hc} = m_h - m_c = (\Delta m_{hc})^{\rm mix} + (\Delta m_{hc})^{\rm ew} \approx \frac{\mu^2}{m_T - m_S} - 160\,\text{MeV}.
\end{align}
The first contribution is due to singlet-triplet mixing and the second contribution due to electroweak loop corrections~\cite{Ibe:2012sx}. The mass difference between the charged state and the lightest state is different in the two scenarios. Neglecting electroweak corrections and assuming small fermion mixing, one finds
\begin{align}
\Delta m_{c\ell} & = m_c - m_\ell \approx m_T - m_S + \frac{\mu^2}{m_T - m_S},\qquad m_S > \mu^2/m_T,\\\nonumber
\Delta m_{c\ell}' & = m_c - m_\ell' \approx m_T + m_S - \frac{\mu^2}{m_T - m_S},\qquad m_S < \mu^2/m_T.
\end{align}
In Fig.~\ref{fig:spectrum}, we show two typical scenarios of dark fermion spectra. For very small singlet-triplet mixing, electroweak corrections dominate the mass splitting $\Delta m_{hc}$ and the charged fermion is the heaviest state (right panel). For larger mixing, $\chi^+$ and $\chi_h$ become degenerate in mass and eventually flip their positions (left panel). The spectra in the scalar and pseudo-scalar scenarios look very similar, unless the mixing is large. The mass hierarchy plays a crucial role for the collider phenomenology of the dark fermions (see Section~\ref{sec:lhc}).

\section{In case of supersymmetry}\label{sec:susy}
\noindent
Our model can be interpreted as the so-called wino-bino scenario in the MSSM with conserved $R$ parity, where the superpartners of the gauge bosons reside around the electroweak scale and the higgsinos, as well as all scalars but the standard-model-like Higgs boson, are much heavier. This scenario is commonly referred to as \emph{split supersymmetry}~\cite{ArkaniHamed:2004fb} with heavy higgsinos. Here we briefly discuss this supersymmetric realization of our Higgs-portal scenario. We adopt the notation for the wino-bino scenario from Ref.~\cite{Nagata:2015pra} and relate it to ours. The phenomenology at energies around the weak scale is described by an effective Lagrangian, 
\begin{align}\label{eq:lag-susy}
\mathcal{L}_{\text{eff}}^{\text{SUSY}} = -\frac{M_1}{2}\widetilde{B}\widetilde{B} - \frac{M_2}{2}\widetilde{W}^A\widetilde{W}^A + C_1^{(5)}\widetilde{B}\widetilde{W}^A H^\dagger T^A H + C_2^{(5)}\widetilde{B}\sigma^{\mu\nu}\widetilde{W}^A W_{\mu\nu}^A + h.c.,
\end{align}
where $\widetilde{W}^A$ and $\widetilde{B}$ are the wino and bino fields, $C_1^{(5)}$ and $C_2^{(5)}$ are Wilson coefficients of the dimension-five operators.~\footnote{We have neglected dimension-5 operators with two winos or two binos, since they are not relevant for our purposes.} The standard-model-like Higgs field $H$ is a linear combination of the two Higgs doublets $H_u$ and $H_d$~\cite{ArkaniHamed:2004fb}. This Lagrangian is to be compared with the Higgs-portal Lagrangian in Eq.~(\ref{eq:lag}). The mass para\-me\-ters $m_S$ and $m_T$ in our model correspond with the gaugino masses $M_1$ and $M_2$. The Higgs-portal coupling $\mu/v$ is related to the higgsino mass parameter $\widetilde{\mu}$ through
\begin{equation}\label{eq:higgsino}
  \frac{\mu}{v} = \frac{\kappa_{ST}}{\sqrt{2}}\frac{v}{\Lambda} = \sin(2\beta) \tan\theta_w \frac{M_W^2}{v|\widetilde{\mu}|},
\end{equation}
where $\tan\beta = v_u/v_d$ is the ratio of vacuum expectation values of the two Higgs fields $H_u$ and $H_d$, and $M_W$ is the mass of the $W$ boson. The cutoff $\Lambda$ can thus be interpreted as the higgsino mass $\widetilde{\mu}$, provided that $\kappa_{ST}$ and $\sin(2\beta)$ are not suppressed. For $\tan\beta = 1$, a portal coupling of $\mu/v = 0.01$ corresponds to a higgsino mass of $\widetilde{\mu} = 1.4\,\text{TeV}$. As we will see in Section~\ref{sec:lhc}, collider searches with long-lived mediators probe parameter regions with $\mu/v \ll 0.01$ and thus higgsino masses well beyond resonant production at the LHC.

The scalar and pseudo-scalar scenarios correspond to different choices of the phases in the diagonalization of the bino-wino mass matrix (see Eq.~(17) in~\cite{Nagata:2015pra}),
\begin{align}\label{eq:SUSY-phases}
scalar\,scenario: & \quad \alpha=\beta=\phi=0,\\\nonumber
pseudo\text{\,-}scalar\,scenario: & \quad \alpha=\beta=0,\,\phi=\pi/2.
\end{align}
Using these relations, one can directly interpret the results of our work in the wino-bino scenario. The pseudo-scalar scenario is a special case of the wino-bino scenario the complex MSSM with flipped bino mass $M_1 \to - M_1$. Signatures of long-lived winos scenario have been discussed in the scalar scenario~\cite{Buckley:2009kv,Rolbiecki:2015gsa,Bramante:2015una}. To the best of our knowledge, a comprehensive analysis of LHC signatures with displaced particles has not been performed yet.  The phenomenology of mediator decays in the pseudo-scalar scenario has been much less explored~\cite{Bharucha:2013epa}. Our work can serve as a framework to systematically search for supersymmetric gauginos in the limit of higgsino decoupling. Notice that in the MSSM the dark matter phenomenology and the resulting collider signatures can be altered by the presence of other light superpartners or additional scalars. The interpretation of our results in a specific scenario should thus be done with care.

\section{Dark matter scattering off atomic nuclei}\label{sec:scattering}
\begin{figure}[!t]
\centering
\includegraphics[height=2.5in]{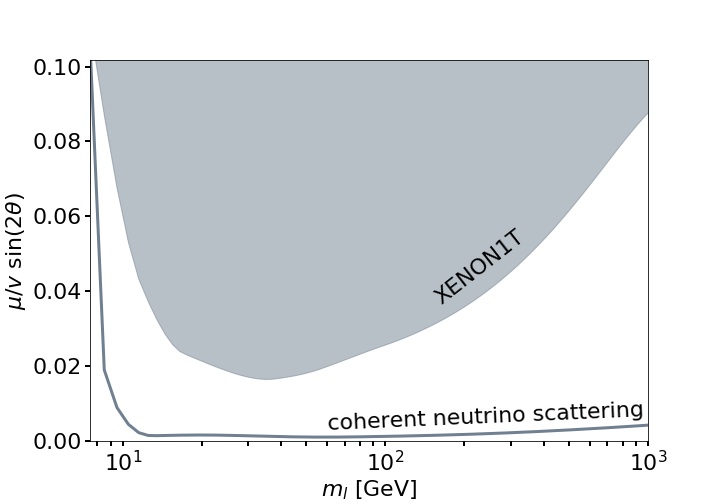} \hspace*{0.5cm}
\caption{\label{fig:direct-detection} Bounds from Xenon1T~\cite{Aprile:2018dbl} on the Higgs coupling to dark matter, as a function of the dark matter mass. The grey region is excluded at $90\,\%$ confidence level. The results have been obtained using \texttt{micrOMEGAs}~\cite{Belanger:2018mqt}.}
\end{figure}

\noindent
From now on, we will interpret the lightest neutral fermion $\chi_\ell$ as a dark matter candidate. An important bound on our model is derived from searches for dark matter scattering off atomic nuclei in direct detection experiments. Since $\chi_\ell$ does not couple to the $Z$ boson, spin-independent scattering is mediated by Higgs boson exchange via a scalar current.\footnote{One-loop contributions to the scattering amplitude scale as $(g\sin\theta)^2/(16\pi^2)$ or smaller and are thus expected to be subdominant.} Thanks to the small momentum transfer, the interaction of dark matter with the quarks inside the nucleon can be described by an effective Lagrangian,
\beq\label{eq:dd-eff}
\mathcal{L}_{\rm eff} = -\,\mathcal{C}_h^q\,(\bar{\chi}_\ell\chi_\ell)\,(\bar{q}q)\,,\qquad \mathcal{C}_h^q = \sqrt{2}\,G_F \frac{m_q}{M_h^2}\,\mu\sin(2\theta),
\eeq
where $G_F$ is the Fermi constant and $M_h$ is the Higgs mass. Notice that the Higgs coupling to two dark matter states is not affected by a chiral rotation and is scalar, regardless of whether the Higgs portal has a scalar or pseudo-scalar structure. In terms of the effective interaction, the coupling of dark matter to protons can be expressed as
\begin{align}
f_p =\sum_{q=u,d,s} f_{Tq}^{(p)} \mathcal{C}_h^q \frac{m_{p}}{m_q} + \frac{2}{27}f_{TG}^{(p)}\sum_{Q=c,b,t} \mathcal{C}_h^Q\frac{m_{p}}{m_Q},\qquad f_{TG}^{(p)} = 1-\sum_{q} f_{Tq}^{(p)},
\end{align}
and analogously for neutrons with $p\to n$. The contribution of a particular quark flavor to the proton mass is $m_p\,f_{Tq}^{(p)} = \langle p|m_q\,\bar q q|p\rangle$, determined experimentally. The cross section for spin-independent dark matter scattering off a nucleus with $Z$ protons and $A$ nucleons in total is given by
\begin{align}
\sigma_A = \frac{4\mu_A^2}{\pi}\big\vert Z f_p+(A-Z) f_n\big\vert ^2 \approx \sigma_n\frac{\mu_A^2}{\mu_n^2}A^2,\qquad \sigma_n=\frac{4\mu_n^2}{\pi}f_n^2,
\end{align}
where $\mu_i^2= m_\chi^2 m_i^2/(m_\chi+m_i)^2$ is the reduced dark matter-nucleon mass and $\sigma_n$ is the cross section for spin-independent dark matter-nucleon scattering. The currently strongest upper bound on nucleon scattering has been obtained by the Xenon1T collaboration~\cite{Aprile:2018dbl},
\beq
\sigma_n < 4.1\times 10^{-47}\,\text{cm}^2\qquad \text{for}\quad m_{\chi} = 30\,\text{GeV}.
\eeq
In our model, this translates to a strong bound on the Higgs coupling to dark matter, which we show in Fig.~(\ref{fig:direct-detection}) as a function of $m_\ell$.
 For instance, for a dark matter mass of $100\,\text{GeV}$ we obtain
\begin{align}
\frac{\mu}{v}\sin(2\theta) \approx \frac{\mu}{v}\frac{2\mu}{m_T - m_S} \lesssim 0.03\qquad \text{for}\quad m_{\ell} = 100\,\text{GeV}.
\end{align}
For sizeable mass splittings $m_T - m_S \gg \mu$, fermion mixing is small. In this regime, the Xenon1T results can be interpreted as an upper bound on the mass splitting between the mediators,
\begin{align}
(\Delta m_{hc})^{\rm mix} \approx \frac{\mu^2}{m_T - m_S} \lesssim 4\,\text{GeV} \qquad \text{for}\quad m_\ell = 100\,\text{GeV}.
\end{align}
The mediator states are thus nearly degenerate in mass. For smaller mass splittings $m_T - m_S \approx 15-30\,\text{GeV}$, as favored by co-annihilation, $\mu/v$ must be suppressed to evade direct detection.~\footnote{Alternatively, direct detection bounds can be evaded in models where spin-independent nucleon scattering is absent by construction~\cite{Gross:2017dan}.} In this regime, fermion mixing can still be close to maximal. As $\mu/v$ is lowered below the Xenon1T bounds, fermion mixing decreases and dark matter decouples from the standard model. Future direct detection experiments are expected to probe even smaller dark matter-nucleon scattering cross sections. If they were able to test rates comparable to coherent neutrino scattering (see Fig.~\ref{fig:direct-detection}), this would probe Higgs couplings at the permil level.

\section{Dark matter annihilation and relic abundance}\label{sec:annihilation}
\noindent
The interpretation of the lightest state as dark matter strongly depends on the Higgs-portal strength $\mu/v$. We consider couplings that are large enough for dark matter to be in thermal equilibrium with the primordial plasma before freeze-out. The relic abundance is then determined by the freeze-out of processes that change the dark matter number density. In Table~\ref{tab:annihilation}, we show the annihilation and scattering processes relevant around the freeze-out temperature and their dependence on the model parameters $\mu/v$ and $\theta$. Throughout our analysis, we consider dark matter masses below $1\,\text{TeV}$, corresponding to the region that can be probed at the LHC. In this mass range, the effect of Sommerfeld enhancement on pair annihilation is mild and will be neglected in our analysis~\cite{Hisano:2006nn}.
\begin{table}[!t]
\centering
\begin{tabular}{|c|l|c||c|l|c|}
\hline
& \ process & scaling & & \ process & scaling \\
\hline
\ pair annihilation \ & \ $\chi_\ell\chi_\ell \to W^+ W^-$ & \ $(g\sin \theta)^4$ & \ mediator annihilation \  & \ $\chi_h\chi_h \to W^+ W^-$ & \ $(g\cos\theta)^2$ \ \\
& \ $\chi_\ell\chi_\ell \to h^\ast \to f\bar{f},VV$ & \ $(\mu\sin(2\theta)/v)^2$ \ & & \ $\chi_h\chi^+ \to  f\bar{f}',VV$ & \ $(g\cos\theta)^2$ \ \\
& \ $\chi_\ell \chi_\ell \to h h$ & \ $(\mu\cos(2\theta)/v)^4$ \ & & \ $\chi^+\chi^- \to  f\bar{f},VV$ & \ $g^2$ \ \\
\hline
\ co-annihilation \ & \ $\chi_\ell\chi^+ \to f\bar{f}',VV$ & \ $(g\sin\theta)^2$ \ & \ mediator decays \ & \ $\chi^+ \to  \chi_\ell f\bar{f}'$ & \ $(g\sin\theta)^2$ \ \\
 & \ $\chi_\ell \chi_h \to W^+ W^-$ & \ $(g\sin\theta)^2$ & & \ $\chi_h \to \chi_\ell f\bar{f}$ & \ $(\mu/v)^2$ \ \\
 & \ $\chi_\ell\chi_h \to h^\ast \to f\bar{f},VV$ & \ $(\mu/v)^2$ \ & & & \\
\hline
\ co-scattering \ & \ $\chi_\ell f \to \chi^+ f'$ & \ $(g\sin\theta)^2$ & \ scattering \ & \ $\chi_\ell f \to \chi_\ell f$ & \ $(\mu\sin\theta/v)^2$ \ \\
& \ $\chi_\ell f \to \chi_h f$ & \ $(\mu/v)^2$ \ & & & \\
\hline
\end{tabular}
\caption{Dark fermion processes around freeze-out and how their rates scale with the Higgs-portal coupling $\mu/v$ and the fermion mixing angle $\theta$. Standard-model fermions and vector bosons are denoted by $f,\,f'$ and $V$. \label{tab:annihilation}}
\end{table}
Scattering off standard-model fermions keeps the dark fermions in thermal equilibrium. Co-scattering, mediator decays and inverse decays ensure chemical equilibrium among the dark fermions.

\subsection{Pair annihilation}
\begin{figure}[!t]
\centering
\includegraphics[height=2.19in]{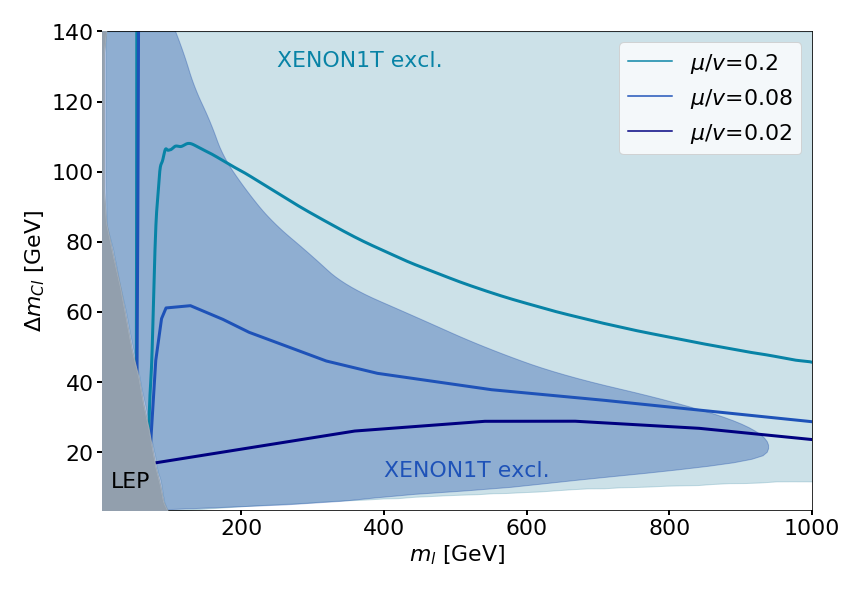} \hspace*{0.2cm} \includegraphics[height=2.19in]{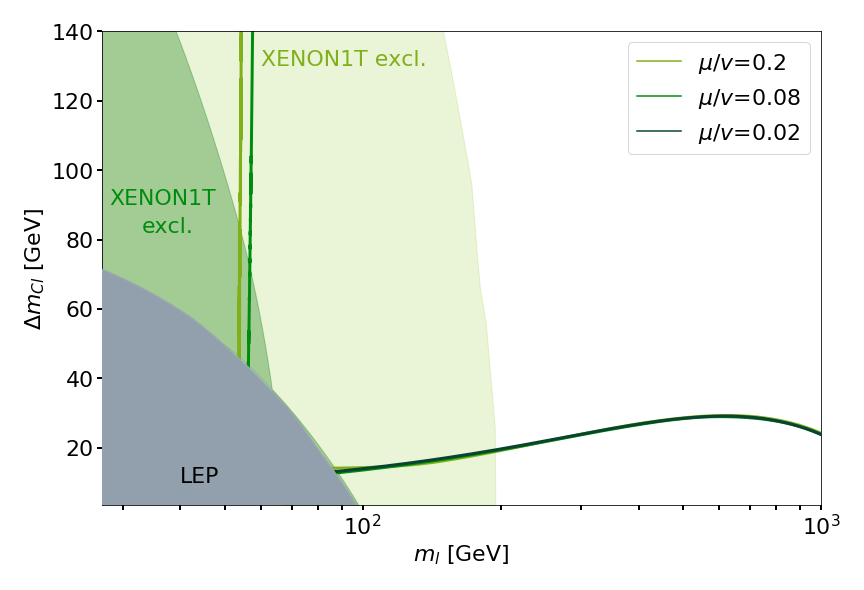}
\caption{\label{fig:relic-abundance}Relic dark matter abundance in the scalar scenario (left) and pseudo-scalar scenario (right). Colored curves show the observed abundance of $\Omega_\chi h^2 = 0.12$ for fixed values of $\mu/v$. Shaded regions are excluded by Xenon1T at $90\,\%$ CL~\cite{Aprile:2018dbl} for the value of $\mu/v$ of the same color. The grey region is excluded by LEP.}
\end{figure}

\noindent Dark matter pairs can annihilate through the Higgs boson or through weak charged currents. Since direct detection results set stringent bounds on the Higgs coupling and also on the mixing of dark fermions below the TeV scale, pair annihilation is suppressed. In Figure~\ref{fig:relic-abundance}, we show the parameter regions of our model that are excluded by Xenon1T in the $\Delta m_{c\ell} - m_\ell$ plane for three values of $\mu/v$. The grey areas are excluded by searches for pair-produced charged fermions at LEP~\cite{Abdallah:2003xe}. For $\mu/v=0.2$, Xenon1T excludes large parts of the parameter space, while for $\mu/v=0.02$ Xenon1T sets weaker bounds than LEP. Notice that in the pseudo-scalar scenario the bounds from direct detection are much weaker, since the mixing $\theta$ for fixed $\mu/v$, $m_\ell$, and $\Delta m_{c\ell}$ is smaller than in the scalar scenario.

The colored curves show the observed relic abundance $\Omega_\chi h^2 = 0.1199$~\cite{Ade:2015xua} for fixed $\mu/v$. To obtain these predictions, we have implemented our model in \texttt{micrOMEGAs}~\cite{Belanger:2018mqt}. The vertical lines around $m_\ell = 63\,\text{GeV}$ correspond to the freeze-out of pair annihilations through the Higgs resonance. As can be seen in the figure, resonant annihilation is compatible with the observed relic abundance even for small $\mu/v$. For larger dark matter masses, the relic abundance is determined by pair annihilation through gauge interactions, provided that fermion mixing is sizeable. In the scalar scenario, this is the case for $\mu/v=0.2$, while in the pseudo-scalar scenario pair annihilation is too small to provide the correct relic abundance.

\subsection{Co-annihilation}\noindent For smaller $\mu/v$, pair annihilation is suppressed by $\theta^4$ and becomes irrelevant around the freeze-out temperature. The relic abundance is now set by co-annihilation processes like $\chi_\ell \chi^+ \to f\bar{f}$~\cite{Edsjo:1997bg}, which scale as $\theta^2$. Since the thermally averaged co-annihilation cross section is Boltzmann-suppressed, a moderate mass difference of $\Delta m_{c\ell}/m_\ell \approx 10\,\%$ is required to prevent overabundance. In the scalar model, for large dark matter masses co-annihilation becomes relevant at larger $\mu/v$, since the pair annihilation rate is smaller than for lighter dark matter. Similarly, in the pseudo-scalar model co-annihilation starts at larger $\mu/v$ than in the scalar model, because the mixing suppression of pair annihilation is stronger.

\subsection{Mediator annihilation}\noindent For even smaller $\mu/v$, the co-annihilation rates become inefficient and dark matter decouples earlier than the heavier dark fermions. The relic density in this case is determined by mediator annihilation $\chi^+\chi^-\to f\bar{f}$ (and similar processes involving $\chi_h$), which can still change the dark matter number density through mediator decays $\chi^+ \to \chi_\ell f \bar{f}'$. Mediator annihilation is not suppressed by fermion mixing (see Table~\ref{tab:annihilation}), but the thermally averaged rate is Boltzmann-suppressed by the mediator mass. Therefore $\chi^+$ and $\chi_h$ should not be much heavier than $\chi_\ell$. Mediator annihilation thus predicts a compressed spectrum, similar to co-annihilation but with smaller dark matter couplings.\\

In summary, direct detection sets an upper bound of $\mu/v \lesssim 0.2\ (0.6)$ on viable thermal relics below 1~TeV in the scalar (pseudo-scalar) scenario. In the pseudo-scalar scenario, all processes decouple for comparably larger Higgs-portal couplings, due to the smaller fermion mixing in this scenario. In particular, for all three displayed values of $\mu/v$ the relic abundance away from the Higgs resonance is set by mediator annihilation.

\subsection{Co-scattering}\noindent In the region of mediator annihilation, dark matter is still in chemical equilibrium with the heavier dark fermions through co-scattering and decays. While decays, co-scattering and co-annihilation all scale as $\theta^2$ or $(\mu/v)^2$, the latter is relatively suppressed by the number density of the non-relativistic mediators and thus decouples earlier. In this regime, determining the relic abundance requires solving the coupled system of Boltzmann equations for the number density evolution of all dark fermions, taking into account co-scattering and decay processes. This approach, however, is not included in automated programs such as \texttt{micrOMEGAs}, \texttt{DarkSUSY} and \texttt{MadDM}~\cite{Belanger:2018mqt,Bringmann:2018lay,Backovic}. While a detailed numerical analysis of the non-equilibrium processes is beyond the scope of our work, we offer a qualitative discussion of the dark matter phenomenology in this regime.

Mediator annihilation determines the relic abundance, as long as mediator decays are prompt around the freeze-out temperature. For a fixed value of $\mu/v$, decays are still rapid in our model when co-annihilation processes have already decoupled. Once the mediator decays drop below the Hubble rate, the dark matter number density can only decrease through co-scattering processes $\chi_\ell\,\text{SM} \leftrightarrow \chi^+\,\text{SM}$ and $\chi_\ell\,\text{SM} \leftrightarrow \chi_h\,\text{SM}$, followed by mediator annihilation. This happens only for very small $\mu/v$, where mediator decays decouple before co-scattering processes. Eventually, the latter decouple as well and dark matter departs from chemical equilibrium, while the mediators remain in equilibrium. The relic abundance is now driven by the freeze-out of co-scattering processes and thus very sensitive to the strength of the Higgs portal. Similar scenarios, dubbed conversion-driven freeze-out, have been identified in Refs.~\cite{DAgnolo:2017dbv,Garny:2017rxs}. In non-standard cosmological scenarios with early matter domination, the observed relic abundance can also be obtained out of thermal equilibrium~\cite{Drees:2018dsj}.\\

\begin{figure}[!t]
\centering
\includegraphics[height=2.4in]{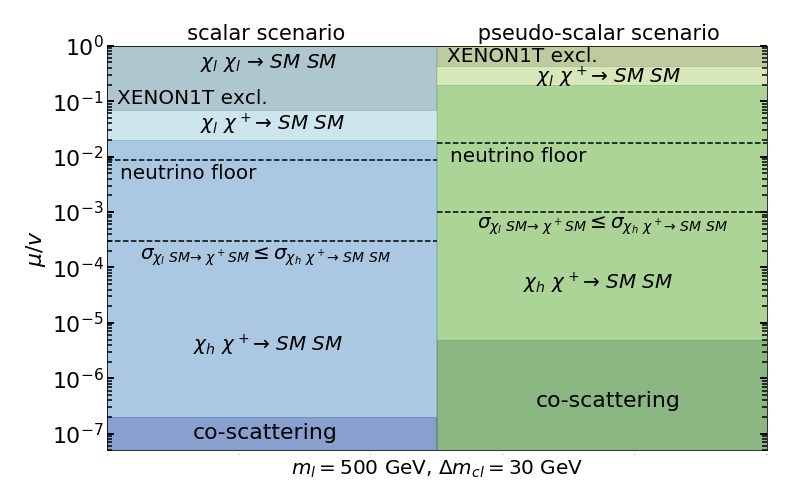}
\caption{\label{fig:relic-regimes}Phases of dark matter freeze-out for small Higgs-portal couplings $\mu/v$. The colored areas denote regions of pair annihilation and co-annihilation, mediator annihilation, and co-scattering, from top to bottom. Displayed processes dominate the relic abundance upon freeze-out. ``SM'' refers to standard-model particles $f,f',V$.}
\end{figure}

In Figure~\ref{fig:relic-regimes}, we illustrate the different phases of dark matter freeze-out for small Higgs-portal couplings in the scalar scenario (blue) and the pseudo-scalar scenario (green).
 For concreteness, we choose a benchmark point with $m_\ell = 500\,\text{GeV}$ and a mass splitting of $\Delta m_{c\ell} = 30\,\text{GeV}$, as it is typical for efficient co-annihilation and mediator annihilation. For other dark matter masses, the main features of the phase diagram are very similar. In the pseudo-scalar scenario, each phase is reached at a larger coupling $\mu/v$. This is due to the fact that the freeze-out of the relevant processes is very sensitive to the fermion mixing, which is smaller than in the scalar scenario~\cite{Baer:2005jq}. The different Lorentz structure of Higgs and gauge couplings in both scenarios (cf. Eq.~(\ref{eq:couplings-small})) has only a subleading effect on the annihilation rates. In particular, all processes that dominate the dark matter abundance can proceed in an $s$-wave in either scenario~(cf. Ref.~\cite{Drees:1992am}).

For large values of $\mu/v$, the relic abundance is set mostly by pair annihilation and co-annihilation. Part of this region is excluded by Xenon1T (shaded in grey). For $\mu/v\lesssim 0.02\ (0.2)$ in the scalar (pseudo-scalar) scenario, the relic abundance is determined by mediator annihilation, mostly by $\chi_h \chi^+ \to \text{SM SM}$. Part of this region might be probed by future direct detection experiments, but the region below the neutrino floor  (indicated by a dashed line) is not accessible with current methods. At $\mu/v\simeq 3\times 10^{-4}\ (10^{-3})$, the thermally averaged co-scattering rate becomes smaller than the (thermally averaged) mediator annihilation rate (see the dashed line). Mediator decays $\chi^+\to\chi_\ell f\bar{f}'$ remain fast, so that co-scattering does not affect the relic abundance yet. At $\mu/v\simeq 2\times 10^{-7}\ (5\times 10^{-6})$, charged mediator decays drop below the Hubble rate and the relic abundance is determined by co-scattering.\footnote{The heavy neutral fermions become long-lived already at larger values of $\mu/v\sim 10^{-4}\ (0.005)$, due to the smaller decay rate for $\chi_h$ (see Section~\ref{sec:lhc}). However, since processes involving $\chi_h$ contribute less than about $10\,\%$ to mediator annihilation in this region, we expect that the relic abundance can still be reliably obtained without including co-scattering in the Boltzmann equations.}

Thermal equilibrium is preserved for couplings well below the range considered in this work. The regimes of mediator annihilation and co-scattering thus provide us with a thermal dark matter candidate that cannot be tested by direct detection experiments. Therefore colliders play an important role in probing Higgs portal dark matter with tiny portal couplings.

\section{Long-lived mediators at the LHC}\label{sec:lhc}
\noindent
The hypothesis of a Higgs-portal dark matter relic with small couplings is directly testable at colliders. In this section, we investigate the LHC phenomenology of our model, constraining the parameter space with existing searches and predicting new observables that test regions that have not been explored yet. At the LHC, the mediators are pair-produced through Drell-Yan-like processes and subsequently decay into the lightest dark fermion. Two examples of such processes are shown in Figure~\ref{fig:production}. The production rate is set by the invariant mass of the mediator pair and the weak gauge coupling.\footnote{For large fermion mixing, the production rate of $pp\to \chi^+\chi_h$ is suppressed by a factor $(g\cos\theta)^2$.} LEP has set a lower bound on the mass of the charged fermion, $m_c \gtrsim 100\,\text{GeV}$. Direct detection results imply that the Higgs-portal coupling must be small, $\mu/v \lesssim 0.2$ for mediators below the TeV scale. Furthermore, direct detection sets an upper bound on the mass splitting between mediators, $\Delta m_{hc}$. Viable scenarios of dark matter freeze-out favor a moderate mass splitting between dark matter and the mediators, $\Delta m_{c\ell}$. These requirements determine the parameter region of interest as
\begin{align}\label{eq:params}
m_c \gtrsim 100\,\text{GeV},\qquad \Delta m_{hc} \lesssim \text{ few GeV},\qquad \Delta m_{c\ell} \approx 15-30\,\text{GeV}, \qquad \mu/v < 0.2.
\end{align}
In this parameter range, fermion mixing is small. At the LHC, we search for a compressed spectrum of dark fermions, featuring potentially long-lived mediators with soft decay products and missing energy. In what follows, we first determine the lifetimes of the mediators and then discuss the resulting signatures and how to test them.
\subsection{Mediator decays}
\noindent
The lifetime of the mediators depends sensitively on the Higgs-portal coupling, $\mu/v$, and on the mass difference between the dark fermions in the initial and final states, $\Delta m$. Neglecting a potential phase-space suppression, the mediator decay width scales as
\begin{align}
\Gamma_{\chi} \sim \left(\frac{\mu}{v}\right)^x(\Delta m)^y,
\end{align}
where $x$ and $y$ depend on the decay process. For decays into the lightest state, $\chi^+ \to \chi_\ell$ and $\chi_h \to \chi_\ell$, the mass splitting is sizeable, $\Delta m_{c\ell} \approx \Delta m_{h\ell} \approx 15-30\,\text{GeV}$. The mediator is thus long-lived on collider scales only if its decay is suppressed by a small portal coupling $\mu/v$. In the decay $\chi^+ \to \chi_h$, in turn, $\Delta m_{hc}$ can be arbitrarily small, as we discussed in Sec.~\ref{sec:model}. In this case, the mediator decay is suppressed by the mass splitting.

Let us first focus on the charged state $\chi^+$. Depending on the Higgs-portal strength, $\chi^+$ can decay either via the two-body decay $\chi^+ \to \chi_h\,\pi^+$ or the three-body decays $\chi^+\to \chi_\ell\,\ell^+\nu$ and $\chi^+\to \chi_\ell+ hadrons$. The two-body decay is kinematically allowed for mass splittings larger than the pion mass. In our model this is fulfilled for very small $\mu/v$, where the mixing-induced mass difference is negligible and $\Delta m_{hc} \approx -160\,\text{MeV}$, see Eq.~(\ref{eq:deltahc}). For $m_\pi < |\Delta m_{hc}| \ll m_c$, the decay width is given by
\begin{align}\label{eq:gammapi}
\Gamma_\pi = \frac{2G_F^2}{\pi} |V_{ud}|^2 f_\pi^2 \cos^2\theta\, |\Delta m_{hc}|^3\Big(1 - \frac{m_\pi^2}{(\Delta m_{hc})^2}\Big)^{1/2}\,,
\end{align}
where  $f_\pi \simeq 130\,\text{MeV}$ is the pion decay constant and $V_{ud}$ is a CKM matrix element. The decay rate is strongly suppressed by the small mass difference $\Delta m_{hc}$, as well as by the limited kinematic phase space. Since fermion mixing is tiny in this regime, $\cos\theta \approx 1$ and the decay rates in the scalar and pseudo-scalar scenarios are the same. The nominal decay length of the charged fermion for $|\Delta m_{hc}| \approx 160\,\text{GeV}$ is given by
\beq\label{eq:chiplus-length}
c\,\tau_{\chi} \approx c/\Gamma_\pi = 7.1\,\text{cm}\times \left(\frac{160\,\text{MeV}}{|\Delta m_{hc}|}\right)^3 \left(1 - 0.76\,\frac{(160\,\text{MeV})^2}{(\Delta m_{hc})^2}\right)^{-1/2},
\eeq
where $\tau_{\chi}$ is the proper lifetime of $\chi^+$. Charged particles with a decay length in the centimeter range leave tracks in the inner layers of the ATLAS and CMS detectors, which we will discuss in more detail below. As $\mu/v$ increases, the mass splitting $\Delta m_{hc}$ drops below the pion mass and the two-body decay is forbidden. In this regime the three-body decays $\chi^+\to \chi_\ell\,\ell^+\nu$ and $\chi^+\to \chi_\ell+ hadrons$ dominate the decay rate. In the scalar and pseudo-scalar scenarios, three-body decays proceed through vector currents (\ref{eq:lag-scalar}) and axial-vector currents (\ref{eq:lag-pseudo-scalar}), respectively. For $\Delta m_{c\ell}\ll M_W,m_c$, the partial width of the leptonic decay $\chi^+\to \chi_\ell\,\ell^+\nu$ is given by~\cite{Djouadi:2001fa}\footnote{In our analysis, we estimate hadronic three-body decays using the branching ratios from Ref.~\cite{Mahbubani:2017gjh}.}
\begin{align}\label{eq:gammaw}
\Gamma_W^S & = \frac{2G_F^2}{15\pi^3}\sin^2\theta(\Delta m_{c\ell})^5 \sim \left(\frac{\mu}{v}\right)^2 \frac{(\Delta m_{c\ell})^3}{v^2},\\\nonumber
\Gamma_W^P & = \frac{2G_F^2}{5\pi^3}\sin^2\theta (\Delta m'_{c\ell})^5 \sim \left(\frac{\mu}{v}\right)^2 \frac{(\Delta m_{c\ell}')^5}{v^2(m_c + m_\ell')^2}
\end{align}
in the scalar and pseudo-scalar scenarios. The mass splitting is very similar in both scenarios, $\Delta m_{c\ell}\approx \Delta m'_{c\ell}$. Contrary to pion decay, three-body decays are not suppressed by the mass splitting, but by the small fermion mixing $\theta$. Notice that in the pseudo-scalar scenario the three-body decays are smaller than in the scalar scenario, leading to a longer mediator lifetime. This is due to the smaller mixing angle, which reduces the weak coupling to dark matter, see Eq.~(\ref{eq:couplings-small}). In the region where three-body decays dominate, the nominal decay length reaches up to $c\tau_{\chi^+}\approx 1.5\,\text{cm}\ (4\,\text{cm})$ in the scalar (pseudo-scalar) scenario. For larger portal couplings, the decay length decreases. Leptonic three-body decays lead to displaced soft leptons, which can in principle be observable at the LHC for displacements larger than $200\,\mu\text{m}$. Hadronic decays lead to soft jets, which are difficult to detect.

The lifetime of the heavy neutral state $\chi_h$ depends on the mass hierarchy of the dark fermions. For sizeable $\mu/v$, corresponding to a normal mass hierarchy (Figure~\ref{fig:spectrum}, left), $\chi_h$ decays dominantly via $\chi_h\to \chi^+\pi^-$, provided that the channel is kinematically allowed. In this case, the decay rate of $\chi_h$ is the same as for $\chi^+$ in the limit $\mu/v\to 0$, given by Eq.~(\ref{eq:gammapi}). For smaller $\mu/v$, the mass hierarchy is inverted (Figure~\ref{fig:spectrum}, right), and $\chi_h$ decays dominantly through $\chi_h \to \chi_\ell\,b\bar{b}$ via an off-shell Higgs boson. For $\Delta m_{h\ell}\ll m_h$, the partial decay width in the respective scenarios is given by
\begin{align}\label{eq:gammah}
\Gamma_h^S & = \frac{G_F}{10\sqrt{2}\pi^3}\frac{m_b^2}{M_h^4} \left(\frac{\mu}{v}\right)^2\cos^2(2\theta)(\Delta m_{h\ell})^5 \sim \left(\frac{\mu}{v}\right)^2 \left(\frac{m_b}{v}\right)^2 \frac{(\Delta m_{h\ell})^5}{v^4},\\\nonumber
\Gamma_h^P & = \frac{3G_F}{280\sqrt{2}\pi^3}\frac{m_b^2}{M_h^4} \left(\frac{\mu}{v}\right)^2\cos^2(2\theta)\frac{(\Delta m_{h\ell})^7}{m_h^2} \sim \left(\frac{\mu}{v}\right)^2 \left(\frac{m_b}{v}\right)^2 \frac{(\Delta m_{h\ell})^7}{v^4m_h^2}.
\end{align}
The decay rate is not suppressed by mixing, but by the small Higgs-portal and bottom Yukawa couplings.\footnote{Loop-induced decays like $\chi_h \to \chi_\ell \gamma$ feature an additional suppression by the small portal coupling and can be neglected~\cite{Nagata:2015pra}.} In the pseudo-scalar scenario, the decay proceeds through a pseudo-scalar current. Parity conservation requires that $\chi_\ell$ and the $b\bar{b}$ pair are emitted in a relative $p$-wave near the kinematic endpoint, resulting in an additional suppression of $(\Delta m_{h\ell}/m_h)^2$. Therefore the heavy neutral fermion lives longer in the pseudo-scalar scenario than in the scalar scenario. Compared with $\chi^+$, the strong suppression of $\chi_h$ decays results in a longer lifetime of the neutral mediator. For $\mu/v\to 0$, $\chi_h$ can be arbitrarily long-lived. The typical signature of a long-lived neutral fermion is a pair of $b$-jets with displaced vertices, which is expected to be detectable by ATLAS and CMS in the range of $1\,\text{cm } < c\tau_{\chi_h} < 1\,\text{m}$~\cite{Nagata:2015pra}.

\subsection{LHC signatures with displaced particles}
\noindent
We are now prepared to investigate the predicted LHC signatures in detail. In Figure~\ref{fig:collider-signals}, we present the parameter region of our model that can be tested with prompt and long-lived mediators in the $m_c$ versus $\mu/v$ plane. For each parameter point, the mass difference $\Delta m_{c\ell}$ is determined by requiring the observed dark matter abundance of $\Omega_\chi h^2 = 0.1199\pm 0.0022$. In most of the parameter space, co-annihilation or mediator annihilation set the relic abundance (cf. Fig.~\ref{fig:relic-regimes}). At the lower end of the plots, the relic abundance starts to be determined by co-scattering, as indicated by a dashed line.

Null results of Xenon1T exclude the upper left corner in the scalar scenario.\footnote{The bound is conservative in the sense that the region is excluded for any dark matter mass below the TeV scale.} In the pseudo-scalar scenario, direct detection bounds are weaker and do not appear in the figure. In either case, current direct detection experiments do not probe parameter regions with long-lived mediators yet. If future experiments became sensitive to scattering rates comparable to coherent neutrino scattering, they would test the region with $\mu/v \gtrsim 0.01$. Indirect detection cannot probe small portal couplings, as dark matter pair annihilation is strongly suppressed. Colliders are thus the only terrestrial instruments to date that can test the hypothesis of dark matter from a tiny Higgs portal.

The various signatures are classified as follows. Green regions are already excluded by existing searches; orange regions correspond to predictions for the LHC with full run-II data (displaced $b$-jet pairs) or for the HL-LHC (disappearing charged tracks); red regions have not been explored yet, but can be probed with new signatures we predict in this work. In what follows, we discuss the signatures one by one, starting from tiny portal couplings and moving upwards in the parameter space.

\subsection{Disappearing charged tracks}\noindent When the portal coupling is tiny, the mass difference between $\chi^+$ and $\chi_h$ is induced radiatively by electroweak corrections and decreases with increasing $\mu/v$. In this parameter region, the spectrum is inverted and $\chi^+$ is the heaviest particle in the dark sector (see Figure~\ref{fig:spectrum}, right). If the splitting is saturated by electroweak corrections, the decay $\chi^+\to \chi_h\,\pi^+$ is kinematically allowed and dominant. The charged mediator decays with a nominal length up to $7\,\text{cm}$ (see Eq.~(\ref{eq:chiplus-length})). It leaves a track in the innermost layers of the detector and decays before reaching the outer tracking layers~\cite{Ibe:2006de,Buckley:2009kv,Mahbubani:2017gjh}. The ATLAS collaboration has performed a dedicated search for supersymmetric winos with similar decay length~\cite{Aaboud:2017mpt}, which is directly applicable to our case. Notice that the analysis assumes that the by-product of chargino production, the neutral wino, is stable at collider scales. In our model, $\chi_h$ decays via $\chi_h \to \chi_\ell b\bar{b}$, which is strongly suppressed by the tiny portal coupling in this regime. This assumption is thus fulfilled by our heavy neutral fermion.

Interpreting the ATLAS results in our scenarios, we find that it excludes mediator masses up to $m_c \approx 460\,(480)\,\text{GeV}$ and portal couplings of $\mu/v \lesssim 10^{-6}\,(10^{-5})$ in the scalar (pseudo-scalar) scenario. Notice that parts of the parameter space with disappearing charged tracks correspond to co-scattering. In this region, the mediator's decay length can deviate from that predicted from co-annihilation. Since co-scattering saturates the relic abundance at smaller couplings $\mu/v$, the mass splitting tends to be larger, but without exceeding $\Delta m_{hc} \approx 160\,\text{MeV}$. The shown results are thus not expected to change much in the co-scattering phase. In particular, we expect the area below the plot to be excluded by disappearing charged track searches as well. At the upper edge of the excluded region, the rates for two- and three-body decays are equal, $\Gamma(\chi^+\to \chi_h\,\pi^+)=\Gamma(\chi^+\to \chi_\ell\,\ell^+\nu) + \Gamma(\chi^+\to \chi_\ell + hadrons)$.  In the pseudo-scalar scenario, this condition is met at larger $\mu/v$, due to the smaller three-body decay rate (see Eq.~(\ref{eq:gammaw})). Above this boundary, three-body decays dominate. In the semi-transparent regions, the branching ratio for $\chi^+\to \chi_h\,\pi^+$ is less than $90\,\%$, which may weaken the bounds derived assuming $100\,\%$ decay into pions.

In Ref.~\cite{Mahbubani:2017gjh}, a dedicated study of the HL-LHC prospects for disappearing charged tracks has been performed for supersymmetric higgsinos. Rescaling the production rate in these predictions for our wino-bino scenario, we estimate that with a luminosity of $3\,\text{ab}^{-1}$ the HL-LHC can extend the search to mediator masses up to $1\,\text{TeV}$.

\subsection{Displaced soft leptons}
\begin{figure}[!t]
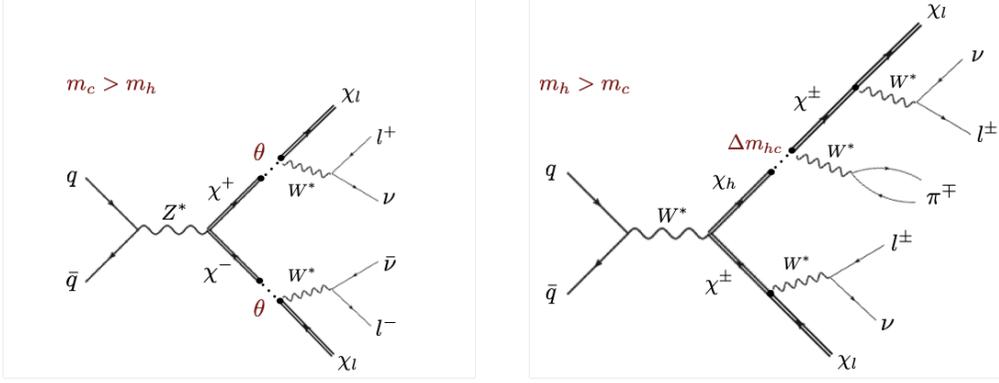

\centering
\includegraphics[height=2.0in]{displaced-leptons.png} \hspace*{0.5cm} \includegraphics[height=2.0in]{displaced+prompt-leptons.png}
\caption{\label{fig:production}New signatures of displaced soft leptons (left) and displaced $+$ prompt soft leptons (right) at the LHC.}
\end{figure}

\noindent In the region where three-body decays dominate the lifetime of $\chi^+$, a soft charged lepton or soft jets are a typical signature of $\chi^+\to \chi_\ell f\bar{f}'$ decays. Here we focus on signatures with displaced soft leptons. The dominant process with a final state of two soft leptons and missing energy is shown in the left panel of Figure~\ref{fig:production}.
 Due to the small mass splitting $15\,\lesssim \Delta m_{c\ell} \lesssim 30\,\text{GeV}$, the transverse momenta of the leptons typically range around $p_{T,\ell} \approx 5-40\,\text{GeV}$. For a sufficiently small portal coupling, the charged mediator can be long-lived and the soft leptons originate from a displaced vertex. The longest decay length of $\chi^+$ is obtained for small portal couplings, $c \tau_{\chi^+} = 1.5\,\text{cm}\ (4\,\text{cm})$ in the scalar (pseudo-scalar) scenario. The smallest decay length that can be observed, $c \tau_{\chi^+} \approx 200~\mu\text{m}$, is limited by the vertex resolution of the detector~\cite{CMS:2016isf}. The corresponding parameter region is shown in Fig.~\ref{fig:collider-signals}.

Searches for displaced leptons have been performed by CMS at $8\,\text{TeV}$ with a lepton momentum cut of $p_{T,\ell} > 25\,\text{GeV}$~\cite{Khachatryan:2014mea} and at $13\,\text{TeV}$ for $p_{T,\ell} > 40\,\text{GeV}$~\cite{CMS:2016isf}. While the momentum cut of the 13-TeV analysis is too strong for our model, a fraction of events with displaced soft leptons falls into the signal region of the 8-TeV analysis. In Ref.~\cite{Bharucha:2018pfu}, this analysis has been recasted for quintuplet dark matter, which leads to the same final state of displaced soft lepton pairs through a decay chain of doubly-charged fermions. We rescale the quintuplet event rates by a factor of $1/4$ and derive the bounds for our model, requiring that the decay length of $\chi^+$ is equal to the decay length of the doubly-charged quintuplet fermion. The resulting $95\,\%$ CL exclusion bound is shown as a green area in Fig.~\ref{fig:collider-signals}. Displaced lepton searches exclude mediator masses up to $m_c \approx 200\,\text{GeV}$ for portal couplings $\mu/v \lesssim 3\times 10^{-5}\ (3\times 10^{-4})$ in the scalar (pseudo-scalar) model. The upper boundary on $\mu/v$ is determined by the minimal decay length that can be probed with the CMS search. The sensitivity of the displaced lepton search extends up to decay lengths of $c\tau_{\chi^+} \gtrsim 7\,\text{cm}$ (and thus to smaller $\mu/v$), where disappearing charged tracks can be observed. In the light green region, the mass splitting $\Delta m_{c\ell}$ is up to $25\,\%$ smaller than in the quintuplet model, and only very few soft leptons still pass the $p_{T,\ell}$ cut. The sensitivity to our model is thus strongly limited in this region.

We therefore suggest to extend searches for displaced lepton pairs to lower transverse momenta. As the lepton momenta spread over a certain range, it could be experimentally beneficial to first lower the threshold for one displaced lepton, i.e., to search for signals with one soft and one harder displaced lepton. Soft displaced leptons can potentially probe much larger mediator masses, as indicated by the red area. The highest accessible masses are experimentally limited by the signal-background discrimination efficiency for small event rates.

\subsection{Displaced $b$-jet pairs}\noindent In the region where $\chi^+$ decays are prompt, parameter regions with heavy mediators can still be probed by displaced signatures if $\chi_h$ is long-lived. This is indeed the case, because the decay $\chi_h \rightarrow \chi_\ell b \bar{b}$ through an off-shell Higgs boson is small (see Eq.~(\ref{eq:gammah})). Produced via $pp\to \chi^+\chi_h$, the slow $\chi_h$ decay leaves a signature of a pair of displaced $b$-jets. Due to the small mass splitting $\Delta m_{h\ell}$, the $b$-jets are rather soft. In Ref.~\cite{Nagata:2015pra}, such a signature has been analyzed in the context of a supersymmetric wino-bino scenario. Projections for the LHC running at $14\,\text{TeV}$ are made, assuming an integrated luminosity of $300\,\text{fb}^{-1}$. The detection criteria for displaced vertices are adapted from an ATLAS study based on $8\,\text{TeV}$ data. Accordingly, a good detection efficiency can be achieved for a decay length in the range $1\,\text{cm} \lesssim c \tau_{\chi_{h}} \lesssim 1\,\text{m}$, reaching its maximum around $c \tau_{\chi_{h}} \approx 10\,\text{cm}$. Reinterpreting these predictions for our scenarios, we derive the parameter region that can be probed with displaced $b$-jet pairs, as indicated by the orange area in Fig.~\ref{fig:collider-signals}. The lower edge of this area corresponds to the largest detectable decay length, $c \tau_{\chi_{h}}= 1\,\text{m}$. The upper edge is set either by the sensitivity limit $c \tau_{\chi_{h}} = 1\,\text{cm}$ or, in the pseudo-scalar scenario for small $m_c$, by $m_h = m_c + m_\pi$, where the spectrum is inverted and $\chi_h$ decays dominantly via $\chi_h \to \chi^+\pi^-$. The hatched regions indicate the (large) uncertainties on the displaced vertex reconstruction~\cite{Nagata:2015pra}. Overall, it is expected that mediator masses up to $m_h \approx 800\,\text{GeV}$ can be probed for appropriate couplings $\mu/v$. Neutral mediators with longer lifetimes, corresponding to a smaller portal coupling $\mu/v$, escape the LHC detector before decaying. Proposed surface detectors might be able to probe this scenario~\cite{Curtin:2018mvb} and thus cover parts of the area below the yellow band in Figure~\ref{fig:collider-signals}.

\subsection{Displaced and prompt soft leptons}\noindent For $m_h > m_c$, the mass splitting due to fermion mixing dominates over electroweak corrections. In this region, the portal coupling is too large to cause displaced $\chi^+$ decays. However, $\chi_h$ can be long-lived for $m_h \gtrsim m_c + m_\pi$, due to a strong phase-space suppression by the small mass difference $\Delta m_{hc}$. From pair-produced mediators via $pp\to \chi_h\chi^+$, one expects one prompt decay, $\chi^+\to \chi_\ell\,\ell^+\nu$, and one slow decay, $\chi_h\to \pi^-/f\bar{f}' (\chi^+ \to \chi_\ell\,\ell^+\nu)$, which produces a displaced soft lepton (Figure~\ref{fig:production}, right). We thus predict a signature with large missing energy and two soft leptons, one displaced and one prompt. The leptons can be either of opposite or of same electric charge, appearing at equal rates. The pion from the decay $\chi_h\to \pi^-\chi^+$ is very soft and not detected.

\begin{figure}[!t]
\centering
\includegraphics[height=2.23in]{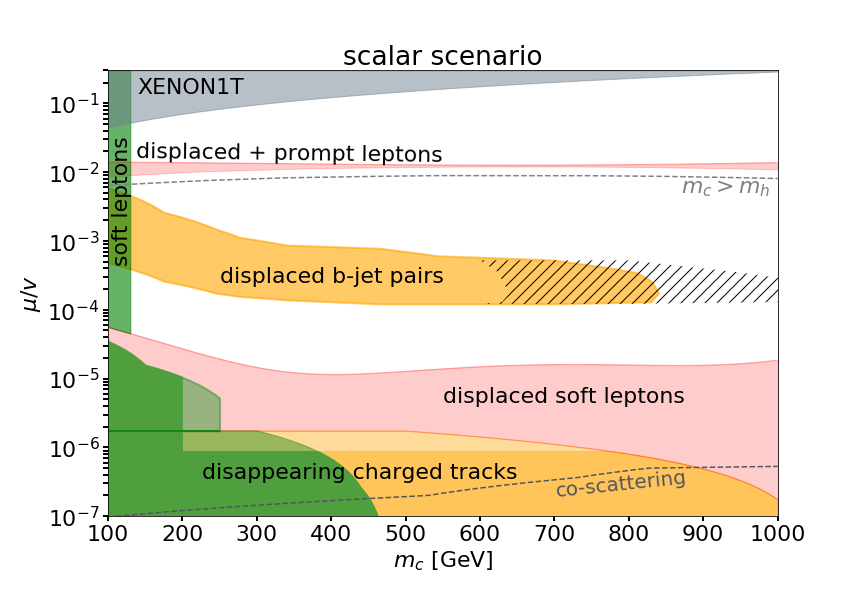} \hspace*{0cm} \includegraphics[height=2.23in]{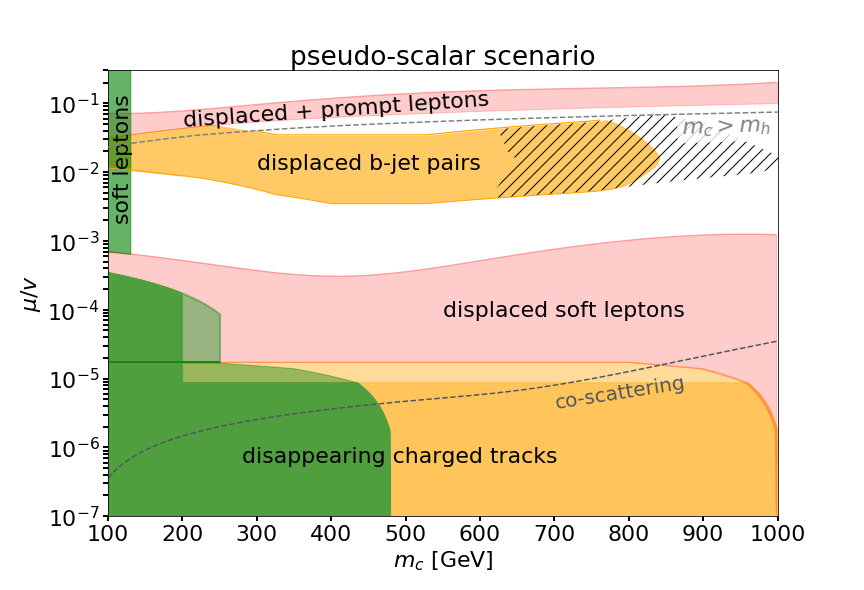}
\caption{\label{fig:collider-signals}LHC signatures of fermion dark matter from a small Higgs portal in the scalar (left) and pseudo-scalar (right) scenarios. Green regions are excluded by existing searches. Yellow regions show projections for the LHC after run II and for the HL-LHC. Red regions feature new signatures that have not been explored yet. The dashed region is due to the unknown detection efficiency of displaced $b$-jets.}
\end{figure}

In Fig.~\ref{fig:collider-signals}, we show the parameter space for displaced and prompt soft lepton signals in red. The upper edge of the area is determined by requiring a minimal nominal decay length of $c\tau_{\chi_h} \gtrsim 200\,\mu\text{m}$. Above this line, the mass splitting $\Delta m_{hc}$ is larger and decays become rapid. The lower edge is set by the kinematic threshold for a two-body decay, $m_h = m_c + m_\pi$. Below this threshold, the heavy neutral fermion decays mostly via $\chi_h \to \chi_\ell b\bar{b}$, and no soft leptons are produced.

\subsection{Prompt soft leptons}\noindent
Mediators with a decay length $c \tau_{\chi^+} < 200~\mu\text{m}$ leave signatures with prompt soft leptons. Both $\chi^+$ and $\chi_h$ decays can produce prompt soft leptons via $\chi^+\to \chi_\ell\,\ell^+\nu$ or $\chi_h \to \pi^-(\chi^+ \to \chi_\ell\,\ell^+\nu)$, respectively. Signatures with two prompt leptons and missing energy are thus expected from $pp\to \chi^+\chi^-$ and $pp\to \chi^+\chi_h$ production. Which process dominates depends on the $\chi_h$ decay branching ratio into leptons. If $\chi_h$ is long-lived, pairs of prompt soft leptons can only be produced via $pp\to \chi^+\chi^-$ with subsequent prompt decays. Searches for pairs of prompt soft leptons have been performed by ATLAS~\cite{Aaboud:2017leg} and CMS~\cite{Sirunyan:2018iwl} in the context of pure wino production via $pp\to \chi^+\chi_h$. The analyses assume a decay via $\chi^+\chi_h \to \chi_\ell W^\ast \chi_\ell Z^\ast \to \chi_\ell \ell^+ \nu\,\chi_\ell \ell^+\ell^-$ to $100\,\%$. However, in our model $\chi_h\to \chi_\ell Z^\ast \to \chi_\ell \ell^+\ell^-$ is loop- and mixing-suppressed and thus small compared to $\chi_h\to \chi_\ell h^\ast \to \chi_\ell b\bar{b}$. In Ref.~\cite{Bharucha:2018pfu}, the CMS analysis~\cite{Sirunyan:2018iwl} has been recasted under the assumption of pure $\chi^+\chi^-$ production for a triplet dark matter model, which corresponds to our scalar scenario. We reinterpret their results and derive a lower bound of $m_c \gtrsim 130\,\text{GeV}$ for $5\times 10^{-5} < \mu/v < 0.02$ in the scalar scenario and $7\times 10^{-4} < \mu/v < 0.07$ in the pseudo-scalar scenario. The excluded parameter region is shown in Figure~\ref{fig:collider-signals} as a green band.

In case of a normal mass hierarchy, $\chi_h$ decays are prompt for $\mu/v\gtrsim 0.02\ (0.07)$ in the scalar (pseudo-scalar) scenario. Soft di-leptons are now produced from both $pp\to \chi^+\chi^-$ and $pp\to \chi^+\chi_h$. Adding the contribution from the latter process enhances our signal rate. However, our $pp\to \chi^+\chi_h$ contribution favors a different kinematic regime and produces only two charged leptons in the final state, compared to three leptons in the experimental analyses. We therefore do not expect a significantly stronger bound than $m_c \gtrsim 130\,\text{GeV}$ when including $\chi^+\chi_h$ contributions. The sensitivity could be enhanced by optimizing the signal region for signatures with exactly two soft leptons in the final state. All in all, however, searches for prompt soft leptons are very limited in their mass reach, due to the small production rates and large backgrounds. Searches for displaced and prompt leptons or displaced $b$-jet pairs are expected to be sensitive to much higher mediator masses.\\

\begin{figure}[!t]
\centering
\includegraphics[height=2.3in]{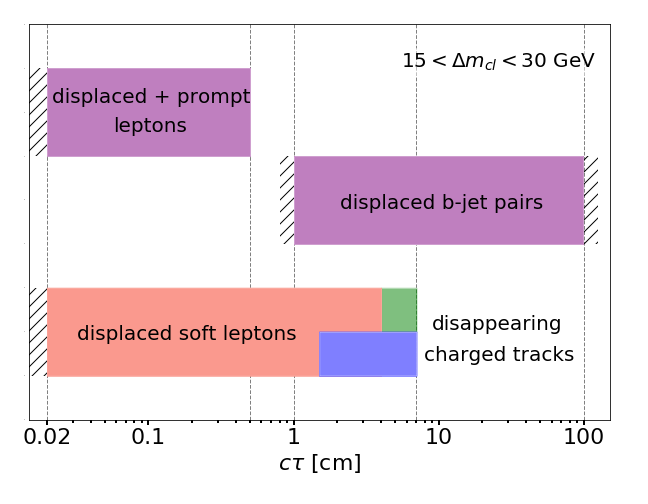}
\caption{\label{fig:ctau}Sensitivity regions for displaced signatures at the LHC. Shown are the ranges of nominal decay lengths of $\chi^+$ (salmon) or $\chi_h$ (purple) that can be tested by ATLAS and CMS. Green and blue parameter regions have already been excluded by searches for disappearing charged tracks. The mass difference $\Delta m_{c\ell}$ is varied within the range that is compatible with the observed dark matter abundance.}
\end{figure}

In Figure~\ref{fig:ctau}, we summarize the various displaced signatures that are predicted from long-lived mediator decays at the LHC. They are classified according to the nominal decay length of the mediator $\chi^+$ (salmon) or $\chi_h$ (purple) in both scenarios. Colored areas correspond to regions of experimental sensitivity with the ATLAS and CMS detectors. Hatched regions could be probed with an extended sensitivity. All other edges are theory bounds, which were explained above for each individual signature. Disappearing charged track searches have already excluded the blue and green parameter regions in the scalar and pseudo-scalar scenario, respectively. It is apparent that our model predicts signatures with different final states in basically all accessible layers of the LHC detectors. The search for dark matter from a small Higgs portal is thus most efficiently done by gathering all these signatures in a combined interpretation.

\section{Conclusions}\label{sec:conclusions}
\noindent
In this work, we have investigated singlet-triplet fermion dark matter interacting with the standard model through a scalar or pseudo-scalar Higgs portal. The nature of the Higgs portal also has implications on the weak charged currents, which are vector-like in the scalar scenario and axial-vector-like in the pseudo-scalar scenario. Both scenarios are related through a chiral rotation of the fermion singlet. Besides the different Lorentz structure of the Higgs and $W$-boson couplings, the fermion mixing in the pseudo-scalar scenario is generically smaller than in the scalar scenario. This leads to different lifetimes of the mediators, with observable consequences for the dark matter and collider phenomenology.

Due to the strong bounds on the portal coupling from direct detection experiments, the thermal relic dark matter abundance relies on co-annihilation with mediators. In the regime of very small portal couplings, co-scattering and mediator decays have a crucial impact on the dark matter number density during freeze-out. For a reliable prediction of the relic abundance, co-scattering and decay processes thus need to be taken into account when solving the coupled system of Boltzmann equations. Since the phase of co-scattering is a common prediction in models with small portal couplings, we suggest to include co-scattering and mediator decays in existing automated tools for relic density calculations.

Thermal dark matter with tiny Higgs couplings and a compressed dark sector implies long-lived mediators. At the LHC, this leads to a plethora of signatures with both prompt and displaced vertices, as well as disappearing tracks. Due to the small electroweak production rates and the softness of the visible decay products, searches for prompt signatures are limited in their mass reach. Current searches for prompt soft leptons cannot probe mediator masses above about $150\,\text{GeV}$. Displaced signatures, in turn, leave clear signatures that can be distinguished from the background with a handful of events. Disappearing charged track searches already exclude mediator masses up to $m_c \approx 460\ (480)\,\text{GeV}$ for tiny portal couplings of $\mu/v \lesssim 10^{-6}\ (10^{-5})$ in the scalar (pseudo-scalar) scenario. Parameter regions with larger couplings can be probed with signatures of displaced $b$-jet pairs, displaced soft lepton pairs, or one displaced and one prompt lepton. Existing searches for displaced leptons probe portal couplings $\mu/v \lesssim 3\times 10^{-5}\ (3\times 10^{-4})$. They exclude mediator masses up to $m_c \approx 200\,\text{GeV}$, but are very limited in their sensitivity, due to the cut on the lepton transverse momentum. Lowering the momentum threshold will strongly enhance the sensitivity of displaced lepton searches to soft decay products, which are typical for compressed dark sectors. Notice that the parameter region with mediator masses not much larger than $m_c \approx 150\,\text{GeV}$ and larger portal couplings can be probed with prompt leptons, as well as displaced and prompt leptons and displaced $b$-jet pairs. A potential discovery of one of these signatures in this region can thus be confirmed by a complementary search for the other signature.

Our analysis shows that mediator masses of several hundred GeV can be accessible at the ATLAS and CMS detectors using the run-II data set. With the data set expected at the HL-LHC, the reach can be extended to probe mediator masses up to the TeV scale and even beyond with future colliders~\cite{Bramante:2015una,Low:2014cba,Mahbubani:2017gjh,Han:2018wus,Kadota:2018lrt}. While neither indirect nor direct detection will be able to test dark matter scenarios with tiny Higgs-portal couplings in the foreseeable future, collider signatures are perfectly characteristic probes. Displaced soft objects in association with missing energy are sensitive to the mass splitting and the coupling of the dark sector at the same time. We enthusiastically encourage the ATLAS and CMS collaborations to exploit the lifetime frontier in the search for Higgs-portal dark matter.

\section{Acknowledgments}
\noindent We thank Felix Br\"ummer, Tao Han, Michel Tytgat and Jos\'e Zurita for helpful discussions and Victor Ananyev for technical support. A warm acknowledgment goes to Nishita Desai for helping us to interpret the displaced lepton searches. We acknowledge support by the DFG Forschergruppe ``New physics at the LHC'' (FOR 2239). AF is funded by the DFG through the research training group ``Particle physics beyond the Standard
Model'' (GRK 1940). SW acknowledges funding by the Carl Zeiss foundation through an endowed junior professorship (\emph{Junior-Stiftungsprofessur}).

\let\oldaddcontentsline\addcontentsline
\renewcommand{\addcontentsline}[3]{}

\let\addcontentsline\oldaddcontentsline

\end{document}